\documentclass[twocolumn,amssymb,aps]{revtex4}

\usepackage{graphicx,dcolumn,bm,amsmath,color,bm}

\newcommand{\eq}[1]{Eq.~(\ref{#1})}

\newcommand{\fig}[1]{Fig.~\ref{#1}}

\definecolor{amber}{rgb}{1.0, 0.75, 0.0}

\newcommand{\eeq}{ \end{equation} }
\newcommand{\beq}{ \begin{equation} }

\newcommand{\eea}{ \end{align} }
\newcommand{\bea}{ \begin{align} }

\newcommand{\bhua}{ {\bf \hat{u}}_{1} }
\newcommand{\bhub}{ {\bf \hat{u}}_{2} }

\newcommand{\bhu}{ {\bf \hat{u}} }

\DeclareMathOperator{\sech}{sech}

\newcommand{\bfr}{ {\bf r} }

\newcommand{\bn}{ {\bf \hat{n}} }
\newcommand{\kbt}{k_{B}T}
\newcommand{\ellp}{\ell^{\prime}}

\newcommand{\red}[1]{ { \color{red} #1 } }
\newcommand{\blue}[1]{ { \color{blue} #1 } }
\newcommand{\amber}[1]{ { \color{amber} #1 } }

\begin{document}
 
\title{Polymeric nematics of associating rods: phase behavior, chiral propagation and elasticity}

\author{H. H. Wensink}
\affiliation{Laboratoire de Physique des Solides - UMR 8502, CNRS, Universit\'e Paris-Sud, Universit\'{e} Paris-Saclay, 91405 Orsay, France}
\email{wensink@lps.u-psud.fr}

\date{\today}

\begin{abstract}
Rod-shaped colloids with attractive tips can form linear aggregates that may subsequently undergo hierarchical self-assembly into nematic fluids.  Inspired by recent modelling efforts on chromonic liquid crystals, composed of discotic building blocks, we formulate a second-virial theory for reversible supramolecular rods. Unlike chromonics, these  systems are capable of forming stable nematic phases in the high-temperature,  monomeric limit in the absence of polymerization. Changing the tip potential from attractive to repulsive thus enables a smooth crossover from a monomeric to a polymeric nematic fluid.  We analyze the isotropic-nematic phase behavior for both regimes and address the nematic elastic properties. The theory accounts for the molecular flexibility and chirality of the filaments and respects the intrinsic chain-length dependence of nematic order. We also discuss the impact of polymerization inhibitors on the phase behavior in the polymeric regime and find that the inhibitors cause a marked narrowing of the isotropic-nematic biphasic region, and generate reentrance nematization as well as a mass density inversion of the coexisting   phases.  We finally discuss the elastic moduli of rod-based polymeric nematics by qualitative comparing their elastic anisotropies with those of chromonic liquid crystals and other nanoparticle-based nematics.
\end{abstract}


\maketitle

\section{Introduction}

Self-assembly through reversible polymerization plays a key role in numerous processes in both passive and living soft matter.  Prominent examples are block-copolymers forming a wide array of micellar structures \cite{riess2003,blanazs2009}.  Microtubules, actin and other filaments found in the cell cytoplasm are composed of dynamically organizing molecular units forming highly interconnected  structures that provide essential mechanical functions to the cell \cite{fuchs1998}.  These filaments can be considered as stiff polymers with a strong directional persistence \cite{gittes1993}.  The latter property my give rise to  hierachical self-assembly into liquid crystalline states, provided correlations between the polymers are strong enough \cite{gennes-prost,plcbook,odijkoverview}.  The simplest of these states is  the nematic fluid which only has short-range positional  order but exhibits long-range orientational ordering of the polymers along a common director. Further complexity can be expected if the polymers are chiral \cite{ho2011}  in which case helical supramolecular organization of the director field may occur such as, in the simplest case, a cholesteric liquid crystal \cite{gray1994,rill1991,dogic-fraden_fil,zanchetta2010a}. 

Filamentous polymers may also emerge from discotic macromolecular mesogens (e.g. dyes) that stack on top of each other due to their strong face-to-face attraction. If the average stack aspect ratio and its number concentration are both sufficiently large, a disorder-order transition may occur from an isotropic fluid of randomly oriented stacks towards a nematic fluid of strongly aligned stacks, referred to as a chromonic liquid crystal \cite{lydon2004,tamchang2008}.   The mechanism underpinning the transition is identical to the one proposed by Onsager for rigid  rods \cite{Onsager}, namely a trade-off between orientational entropy favoring orientational disorder and excluded-volume entropy driving  elongated, rod-shaped objects to align \cite{kuriabova2010,nguyen2014}.  A key attribute of these polymeric systems  is the inherent length-dispersity where  temperature dictates the propensity of the monomeric units to bond together. The distribution of aggregate lengths is further controlled by the degree of filament flexibility, and the overall monomer concentration. The thermodynamics of linear aggregates can be properly captured within a simple second-virial theory, as formulated for e.g. wormlike micelles \cite{vdschoot1994la}. This approach provides a comprehensive route  to predicting the role of each of these control parameters on the aggregation statistics and fluid phase behavior from considering  (local) pair-interactions only  \cite{vdschoot1994epl}. Recent modelling efforts based on similar theoretical concepts have been directed towards understanding the phase behavior and mesochiral properties of DNA oligomers \cite{demichele2012,demichele2016}, the elastic properties of chromonics \cite{romani2018} and the tactoid morphology of amyloid fibrils \cite{bagnani2019}.
   
Most models considered to date share the characteristic feature that the monomeric building blocks are low-anisotropy nanoparticles such as spheres \cite{lu2004,nguyen2018}, short spherocylinders \cite{nguyen2014} or discotic particles \cite{romani2018} each carrying some patchy, associative functionality.  In monomeric form, the shape anisotropy does not permit the formation of liquid crystals.  Elongated rods capable of associating tip-to-tip would be an exception to this case. Recent advances in colloidal self-assembly have opened up ways to fabricate a wide array of complex compound colloidal particles with tunable shape and site-specific interactions \cite{sacanna2013,edwards2007}. The intricate ``supra-colloidal" self-assembly of these patchy and non-isotropic colloids has been pursued in  detail in computer simulation \cite{zhang2004,vananders2014} and is only beginning to be explored experimentally.  A prominent example of distinctly rod-shaped colloids are filamentous {\em fd} virus rods \cite{dogic2016} that have recently been demonstrated to be amenable to tip-functionalization \cite{delacotte2017,repula2019} rendering the inter-rod interactions even more complicated.  In this way, the rods could be made ``polymerizable" by equipping each tip with an attractive site \cite{repula2019}. Contrary to chromonics, the monomeric rods are considered sufficiently  slender to exhibit various types of liquid-crystalline self-organization, including cholesteric order \cite{dogic2000,dogic-fraden_fil}. From a theoretical standpoint, self-associating rods are an appealing model system as their strongly elongated shape facilitates ordering effects at very low filling fractions which justifies the use of a second-virial theory \cite{Onsager}. For the other polymerizing non-isotropic nano-objects considered to date,  the second-virial approximation should only be applicable to the very longest aggregates whose length-to-width ratio tends to infinity. Correlations between the mono- and oligomeric species, on the other hand,  are characterized by non-negligible steric end-effects that requires inclusion of higher-order virials \cite{Onsager,Vroege92,odijkoverview}. A further complication in self-assembly driven nematic order is that the nematic order parameter varies according to the aggregate length, with short aggregates being less strongly ordered than long ones \cite{odijkoverview,vroege1993}. Although this basic relationship can, in fair approximation, be neglected for semiflexible rods \cite{vdschoot1994epl} it should not be overlooked for rigid \cite{wensink2003,Wensink_2019} or weakly flexible rod-shaped mesogens (such as {\em fd})  for which the persistence length amounts to several times the contour length. 
   
In this paper we explore the thermodynamics of linear polymerizing weakly flexible rods using a simple second-virial theory that captures the principal entropic features due to volume  exclusion,  association and the orientational entropy as well as the effect chiral interactions between the rod monomers.  We pay particular attention to the length-dependence of the nematic order parameter. Our theoretical approach is closely related to  the one outlined in Kuriabova et al. \cite{kuriabova2010} where this coupling has been accounted for using a series expansion of the orientation-dependent free energy contributions combined with a trial function for the aggregate distribution \cite{vdschoot1994epl}. Here, we shall use a simpler approach based on Gaussian trial function Ansatz for the orientational probability of the aggregates which enables the thermodynamic expressions to be cast in algebraic form provided the nematic alignment of each aggregates is sufficiently strong. This condition is naturally met in case the monomers are slender rods with sufficient stiffness to enable them to  order nematically.  Our theory allows a facile computation of the phase diagrams as well as the Frank elastic moduli for which analytical expressions for asymptotically strong nematic order limit had already been derived bij Odijk \cite{odijkelastic}.  The inherent length-dispersity of the aggregates is taken into account using the thermodynamically consistent distribution in the nematic phase while the aggregate flexibility is approximated by assuming the aggregate to behave as a collection of rigid filaments with a length set by  the deflection length, which is the principal length scale of a flexible polymer residing in a nematic matrix \cite{romani2018}.

We distinguish two nematic regimes; a polymer-dominated regime at low (effective) temperature  dominated by strong end-to-end aggregation, and a monomeric one  at high temperature where the nematic fluid mostly consists  of mono- and oligomeric rods. The presence of polymerization inhibitors  that irreversibly attach to the aggregate tip thereby preventing further growth will be accounted for using a simple free energy penalty term. It basically forces the system to generate more free ends if the inhibitor concentration exceeds the concentration of aggregate ends. The presence of  inhibitors is shown to strongly affect the isotropic-nematic phase behavior in the  low-temperature regime by generating marked reentrance effects along the  isotropic-nematic binodals as well as a subtle length-dispersity driven mass-density inversion of the coexisting phases. The elastic properties of the polymeric nematic expected at low temperature are drastically different from a high-temperature monomeric nematic fluid. In particular, we find that a polymeric nematic fluid exhibits a much higher splay elasticity but a lower bend modulus than its monomeric counterpart. Also, we observe an exponential increase in the pitch  upon lowering temperature triggered by a strong boost in the twist modulus as the average aggregate length grows larger.  

The  paper is organized as follows. In Section II, we begin by presenting a second-virial free energy and discuss the main entropic and enthalpic contributions underlying the association and ordering processes. In Section III and IV we demonstrate how the entropic terms can be specified from microscopic considerations by employing asymptotic expansions of the angular averages involved. The equilibrium aggregate distribution in isotropic and nematic fluid phases are discussed in Section V and  a simple model mimicking the effect of polymerization inhibitors is discussed in Section VI. Results for the fluid phase behavior are shown in Section VII  and the consequences of polymerization for the  elastic moduli are discussed in Section  VIII. The main conclusions from this work will be formulated in Section IX.

\begin{figure}
\begin{center}
\includegraphics[width=   \columnwidth]{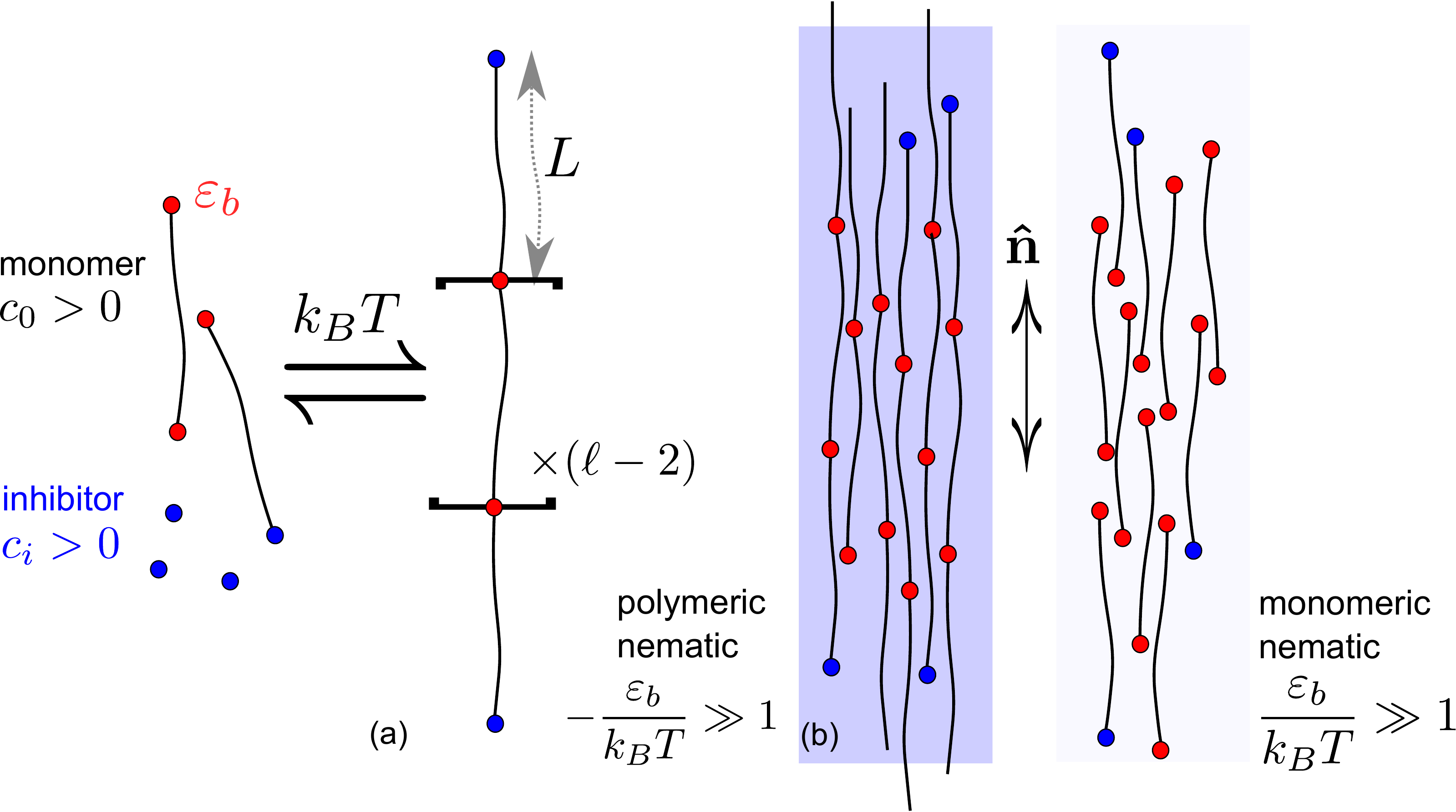}
\caption{ \label{fig1} (a) Filamentous, weakly flexible monomers with contour length $L$ equipped with reactive rod tips (red dots) showing reversible polymerization resulting in a linear aggregate with with aggregation number $\ell $. The growth of the polymer is terminated by the presence of  inhibitors (blue dots) irreversibly binding to the filament tips. (b) At elevated monomer concentration $c_{0}$ the filaments undergo hierarchical self-assembly forming a nematic.   When the rod tip are  attractive ($-\varepsilon_{b}/\kbt  \gg 1$) a length-disperse polymeric nematic fluid is formed. No polymerization occurs for strongly repulsive tips ($\varepsilon_{b}/\kbt  \gg 1$) thus giving a monomeric nematic. }
\end{center}
\end{figure}

\section{Theory for Reversibly Polymerizing Weakly Flexible Chiral Rods}
We start with formulating the free energy per unit volume $V$ of a polydisperse assembly of reversibly polymerizing  rods with  length $L$ and diameter $D \ll L$ and define $\rho(\ell, \Omega)$ as the number density of monomers  aggregated into a slightly flexible polymer with contour length $\ell L$  and orientation described by a  solid angle $\Omega$. We assume that the filaments do not form multi-armed or ring-shaped aggregates and always retain fluid order such that the monomer density is uniform throughout space.  The aggregation number is specified by the integer $\ell =1,2,3,...$.  Following \cite{kuriabova2010}, we write the free energy per unit volume of such an assembly as follows:
 \begin{align}
& \frac{  F}{ V}  \sim \kbt \sum_{\ell} \int  d \Omega   \left [ \ln \left ( v_{0}  \rho (\ell, \Omega) \ell^{-1}  \right ) - 1 \right ] \ell^{-1} \rho (\ell, \Omega) \nonumber \\ 
&   -\varepsilon_{b}  \sum_{\ell} \int d \Omega \ell^{-1} \rho(\ell, \Omega) (\ell -1 ) -  T \sum_{\ell} \int  d \Omega \rho(\ell, \Omega)  S_{\rm or}  \nonumber \\ 
&  + K_{0} + q K_{t} + \frac{1}{2} q^{2} K_{2} 
 \label{free}
\end{align}
where $  \kbt $ denotes the thermal energy in terms of Boltzmann's constant $k_{B}$ and temperature $T$. Furthermore, $v_{0} = (\pi/4)L^{2}D$ is the monomer excluded volume at random (isotropic) orientations. 
The total monomer concentration $\rho_{0}$ is a conserved quantity which requires that the distribution be normalized according to $\rho_{0} = \sum_{\ell} \int  d \Omega \rho(\ell, \Omega )$. 
The first term is related to the ideal gas or mixing entropy containing the ideal translation and orientational entropy of each chain as a whole.
 We have ignored contributions related to the thermal volume of the  monomers that are immaterial for the thermodynamic analysis presented here. 
The second contribution in \eq{free} represents an association energy that drives polymerization. Here, we assume each rod to be equipped with identical attractive patches at either tip such that each rod end can form a single bond  with an adjacent rod thus forming a linear aggregate (see \fig{fig1}).   The free energy per unit volume arising from the polymerized rod segments follows  from the adhesion potential $\varepsilon_{b} <0$ between two adjacent rod segments and the number density  $\rho_{\rm a}(\ell, \Omega) = (1/\ell) \rho(\ell, \Omega)$ of polymers with aggregation number $\ell  $  each containing  $\ell -1$ bonds. At zero temperature, the  energy would be minimal if all monomers  were to join together  forming a single,  infinitely long polymer. Naturally, for $-\varepsilon_{b}$ of the order of the thermal energy $k_{B}T$, the single chain configuration is entropically highly unfavorable in view of the mixing entropy term that  favors a broad distribution of polymers with strongly disperse contour lengths.  
All polymers, irrespective of their contour length, are assumed to be wormlike chains characterized  by a persistence length $\ell_p $ that we express implicitly in units of the monomer length $L$. We assume the monomers to be weakly flexible so that $\ell_{p} > L$, the example in point being {\em fd} virus rods for which $\ell_{p}$ is a few times  $L$ (see  \cite{wang2006,grelet2014} and references therein). We further assume that the inter-filament bonds (indicated by red dots in \fig{fig1}) do not change the intrinsic persistence length of each chain. An alternative model that we will not further consider here is to assume the associating rods to bond at a fixed bond angle $\vartheta$ (assumed small) thus forming 
some kind of freely rotating chain with a  persistence length set by $\ell_{p} = \ell / \ln ( \cos \vartheta )$. The nematic ordering properties of polymers of this kind has been theoretically analyzed  in Refs. \cite{Wessels_2003,Wessels_2006} for the case of non-associating, fixed-length polymers composed of rigid segments with adjacent units experiencing a simple bending potential. These polymers then resemble wormlike chains only when the number of segments is very large. 

Chain flexibility generates an additional entropy $S_{\rm or}$ associated with the internal orientational fluctuations within each polymer of aggregation number $\ell$. This contribution can be specified in the asymptotic limit of strong nematic order.  Although very long  aggregates with contour length exceeding the persistence length are strictly {\em semi-flexible} we shall ignore the possibility of chain backfolding and hairpin configurations that may become prominent in dilute nematics \cite{milchev2018,popadic2018}. 

The last three contributions in \eq{free} capture the correlations between the polymers in a twisted nematic director field.  The degree of helical organization is defined in terms of a helical wave number $q = 2 \pi / p_{c}$ with pitch $p_{c}$ (in units monomer length $L$). In the limit of weak twist considered here, the pitch is assumed to be much larger than the average polymer contour length $ p_{c} \gg \langle \ell \rangle$.  Under these restrictions $q \ll 1$ and the director free energy density takes a simple quadratic form in \eq{free}.  For the achiral nematic phase ($q=0$) only $K_{0}$ remains which is associated with the excluded volume entropy. It reads:
\beq
K_{0} =  \frac{1}{2} \kbt  \sum_{\ell, \ell^{\prime}}  \iint d  \Omega  d \Omega^{\prime} \rho_{l}(\ell, \Omega)  \rho_{l^{\prime}}(\ell^{\prime}, \Omega^{\prime})  2 l^{2} D | \sin \gamma |  
\label{k0pre}
\eeq
The excluded-volume between a pair of wormlike chains is described by considering a chain of  rigid, hard rod segments of  length $l <L$ (with $l  \gg  D$) \cite{Vroege92}. Identifying  the concentration of effective segments  $ \rho_{l}(\ell, \Omega)  = \rho(\ell, \Omega) (L/l)$ one finds that the excluded volume entropy is independent of $l$:
\beq
K_{0} =  \frac{1}{2} \kbt  \sum_{\ell, \ell^{\prime}}  \iint d  \Omega  d \Omega^{\prime} \rho(\ell, \Omega)  \rho(\ell^{\prime}, \Omega^{\prime})  2 L^{2} D  | \sin \gamma |  
\label{k0}
\eeq
The remaining two contributions in \eq{free}  denote the helical amplitude $K_{t}$ and twist elastic modulus  $K_{2}$.   A rigorous statistical mechanical analysis of the elastic moduli of (semi-)flexible polymers is a complex problem requiring a proper description of the  spatial conformations of a flexible rod coupled to a non-uniform director field.  In this paper, we shall  use a simple rescaling of the excluded-volume integrations  in terms of the the principal length-scale  governing orientational correlations of a flexible polymer embedded in a nematic matrix as we will demonstrate shortly.

In order to simplify the analysis of the free energy minimization we  factorize the monomer density using a simple Gaussian trial function describing the orientational probability of the monomers \cite{odijkgauss,odijkoverview}:
\beq
\rho (\ell, \Omega) \sim  \rho_{\ell} \psi_{G} (\theta, \ell)
\eeq
Although the twist of the director field changes the local uniaxial alignment in favour of biaxial order \cite{harris-rmp}, the biaxial perturbation should be very weak for $q \ll 1$ and we  assume  local uniaxial nematic order to be unperturbed.  Then,  the orientational distribution depends solely on the polar angle $\theta$  between the main particle orientation vector $\bhu$ and the nematic director $\bn$ via $\cos \theta = \bhu \cdot \bn$. The Gaussian trial function is given by:
\beq
 \psi_{G} (\theta, \ell) \sim \frac{\alpha_{\ell}}{4 \pi}  \exp \left (  -\frac{1}{2} \alpha_{\ell} \theta^{2}  \right ) 
 \label{gauss}
 \eeq
supplemented with its polar mirror form $\psi (\pi - \theta , \ell)$ along $-\bn$ in order to guarantee local orientational order to be apolar.  The distribution is normalized according to $\int_{0}^{\infty}  d \theta \theta  \int_{0}^{2 \pi}  d  \varphi \psi_{G} (\theta, \ell) \sim 1$ with $\varphi$ indicating the azimuthal angle. The variational parameter  $\alpha_{ \ell} $ is required to be much larger than unity and is expected to grow uniformly with the aggregation number $\ell$. The Gaussian approximation cannot represent isotropic order since, upon taking $\alpha_{\ell} \downarrow 0$, the expression above reduces to zero rather than giving the desired form $\psi = 1/4\pi$.   There are consistent algebraic trial functions for $\psi$  that do correctly render isotropic order in this limit, but these involve more complication distributions that tend to heavily compromize the tractability of the theory \cite{Onsager, francomelgar2008, demichele2016}. 

\section{Asymptotic results for the orientational and excluded volume entropy} 

The factorization Ansatz enables us to greatly simplify the free energy \eq{free} in view of the asymptotic results that are available for the angular averages involved in the various entropic contributions. Introducing brackets to denote Gaussian averages $ \langle \cdot \rangle =  \int_{0}^{\infty}  d \theta \theta  \int_{0}^{2 \pi}  d  \varphi $ we obtain for the ideal orientational entropy of the aggregates:
\begin{align}
\langle \psi_{G} \ln [4 \pi \psi_{G}] \rangle  \sim \ln \alpha_{\ell}  -1
\end{align}
A Gaussian estimate of the orientation entropy arising from aggregate flexibility (of arbitrary strength) is given by  \cite{odijkoverview, Vroege92}:
\begin{align}
 -\frac{S_{or}}{k_{B}} &\sim \frac{\ell }{6 \ell_{p}} (\alpha_{\ell} -1) + \frac{5}{12} \ln  \left [ \cosh \left ( \frac{\ell }{5 \ell_{p}} (\alpha_{\ell} -1 )  \right ) \right ]  \nonumber \\
 & \approx \frac{\ell }{4 \ell_{p}} \alpha_{\ell} + \text{cst}
\end{align}
The linear increase with $\alpha_{\ell} \gg 1$ reflects the considerable loss of conformational entropy a flexible chain undergoes when embedded in a strongly aligned nematic phase.

The excluded-volume term \eq{k0} follows from a double Gaussian average of the sine of the enclosed angle between two rods. This gives up to leading order
\begin{align}
 \langle \langle \psi_{G} \psi^{\prime}_{G}  M_{0}   \rangle \rangle  \sim 2 L^{2} D \left( \frac{\pi}{2} \right ) ^{\frac{1}{2}} \left ( \frac{1}{\alpha_{\ell}} +  \frac{1}{\alpha_{\ellp}} \right )^{\frac{1}{2}} 
\end{align}
using shorthand notation $\psi_{G}^{\prime} = \psi_{G} (\theta^{\prime}, \ellp)$. Although the excluded-volume part of  free energy does not explicitly depend on the polymer contour length, it does so implicitly through $\alpha_{\ell}$. The monotonic increase of $\alpha_{\ell}$ with $\ell$ (as  will be demonstrated shortly) reflects the fact that  long polymers are more strongly aligned than short ones.

\section{Asymptotic results for the helical amplitude and elastic moduli} 

We will now focus on the collective properties of reversibly polymerizing rods  by quantifying the fluctuations and twist of the nematic director field.   For (semi-)flexible polymers, the principal length-scale governing orientational correlations of the polymer embedded in a nematic matrix is the {\em deflection length}  defined as $ \lambda_{\ell} \equiv \ell_{p}/ \alpha_{\ell} $. The deflection length (hereafter expressed in units monomer length $L$) is much shorter than the persistence length and corresponds to the orientational correlation length of a polymer that is strongly confined in a narrow `tube'  generated by the aligned neighboring polymers \cite{odijkoverview,Vroege92}. 
In view of  the strongly polydisperse nature of the nematic phase, the average ratio of the contour length with respect to the deflection length can be estimated from the following expression:
\beq
\left \langle \frac{\ell}{\lambda_{\ell}} \right \rangle = \frac{  \sum_{ \ell} \rho_{\ell}  \alpha_{\ell} \ell_{p}^{-1}}{ \sum_{  \ell} \rho_{\ell} \ell^{-1}} > 1 
\eeq
For all cases discussed in this work,  the contour length will be much larger than the deflection length so that the above condition holds throughout.  Approximate microscopic expressions for $K_{t}$ and $K_{2}$ that feature in the free energy \eq{free} of a nematic fluid of flexible rods can be proposed by assuming each polymer with aggregation number $\ell$ to be composed of rigid sections of length $\lambda_{\ell} < \ell $ with probability density $ \rho_{\ell} (\ell / \lambda_{\ell})$ \cite{romani2018}. Then, following Straley \cite{straleychiral} we write:
\begin{align}
K_{t} & \sim   \frac{1}{2} \sum_{  \ell,   \ellp}   \left ( \frac{\ell \ellp }{\lambda_{\ell} \lambda_{\ellp} } \right )  \rho_{\ell}  \rho_{\ellp} \langle \langle \psi_{G} \dot{\psi}_{G}^{\prime} \Omega^{\prime}_{\perp} M_{t}^{(\ell \ellp)} \rangle  \rangle  \nonumber \\
K_{2} & \sim  \frac{1}{2} \sum_{   \ell  , \ellp}   \left ( \frac{\ell \ellp }{\lambda_{\ell} \lambda_{\ellp} } \right )  \rho_{\ell}  \rho_{\ellp}  \langle \langle \dot{\psi}_{G} \dot{\psi}_{G}^{\prime}  \Omega_{\perp}  \Omega^{\prime}_{\perp}  M_{2}^{(\ell \ellp)}  \rangle \rangle
\label{kaas} 
\end{align}
These expressions involve  the angular derivative of the orientational probability distribution via  $\dot{\psi}_{G} =  \partial \psi_{G} / \partial \cos \theta $. Furthermore, $\Omega_{\perp}$ denotes the component of the rod orientation perpendicular to the local nematic director and the pitch axis.  The kernels describe the interactions between  chiral monomeric rods which we assume to consist of a weak soft potential $u_{c}$  imparting chirality and an achiral hard-core repulsion  generating twist elasticity. The helical amplitude  $M_{t}$ is defined as an integrated (van der Waals) potential \cite{varga2006a,wensinkjackson}. It can be expressed as follows:   
\beq
M_{t}^{(\ell \ellp)} \sim  \int_{\notin v_{\rm excl}} d \bfr \bfr^{\parallel}   u_{c} (\bfr, \Omega, \Omega^{\prime} )
\label{mt}
\eeq
Here, $u_{c}^{(\ell \ellp)}$ denotes some (effective) chiral potential that we will specify later, and $\bfr_{\parallel}$ the component of the centre-of-mass distance between a rod pair along the pitch axis. The second kernel $M_{2}$ describes a generalized excluded-volume between effective achiral  rod monomers of length $\lambda_{\ell}$ and $\lambda_{\ellp}$ and reads \cite{odijkchiral,wensinkjackson}:
\beq
M_{2} ^{(\ell \ellp)} \sim - \kbt \frac{1}{6} \lambda_{\ell} \lambda_{\ellp}  L^{4}  D  | \sin \gamma | (\lambda_{\ell}^{2}  \Omega_{\parallel}^{2} + \lambda_{\ellp}^{2}  \Omega^{\prime 2}_{\parallel}  )
\label{m2}
\eeq
with  $\Omega_{\parallel}$ the rod orientation projected along the pitch axis. 

Let us consider a simple  chiral potential acting between two freely rotating rods  expressed in terms of a commonly used pseudo-scalar form \cite{goossens,vargachiral}:
\beq
u_{c} (\bfr, \Omega, \Omega^{\prime} ) \sim \varepsilon g \left ( r \right ) ( \bhua \times \bhub \cdot {\hat{\bf r}} )   
\eeq
with $g(r)$ some rapidly decaying function of interrod distance and $\varepsilon$ defining the microscopic chiral strength between the particles. We may work out the integrated chiral potential $M_{t}$ corresponding to this potential,  first by defining the pitch axis of the cholesteric to align along the $z-$axis of a Cartesian laboratory frame and define a rod orientation  $\bhu = ( \cos \theta, \sin \theta \sin \varphi, \sin \theta \cos \varphi ) $  in terms of polar and azimuthal angles $(\theta, \varphi)$ with respect to a reference nematic director pointing along the $x-$axis. Then, $\Omega_{\perp}  = u_{y}$  and we  perform a Taylor expansion for $\theta \ll 1$  keeping only the leading order contribution. Some algebraic manipulations along the lines proposed in Refs. \cite{odijkelastic, odijkchiral}  lead to the following asymptotic expression: 
\beq
 \Omega^{\prime}_{\perp} M_{t}^{(\ell \ellp)} \sim   \bar{\epsilon}(LD)^{2} \lambda_{\ell}   \lambda_{\ellp}  \left [ (\theta^{\prime 2} - \theta^{2} ) + |\gamma | ^{2}  \right ] 
 \label{ommt}
\eeq
with $ \bar{\epsilon}$ a dimensionless molecular chiral amplitude combining various microscopic features:
\beq
\bar{\epsilon} \sim \varepsilon  \int_{1}^{\infty} d \bar {r}  \bar{r} g\left ( \bar{r}  \right ) 
\eeq
with $\bar{r} = r/D$. A similar analysis can be performed for the kernel $M_{2}$ underpinning the twist elastic modulus. This reveals the following angular dependence \cite{odijkelastic}:
\begin{align}
&  \Omega_{\perp} \Omega^{\prime}_{\perp} M_{2}^{(\ell \ellp)}   \sim   -  \kbt \frac{\lambda_{\ell} \lambda_{\ellp} L^{4} D}{96} \nonumber \\
 & \times  \left [ | \gamma | (\theta^{\prime 2} + \theta^{2} )(\lambda_{\ellp}^{2} \theta^{\prime 2} + \lambda_{\ell}^{2} \theta^{2})   - |\gamma | ^{3} ( \lambda_{\ellp}^{2} \theta^{\prime 2} + \lambda_{\ell}^{2}  \theta^{2}) \right ]
 \label{omomm2}
\end{align}
We may now perform Gaussian orientational averages of these quantities as per \eq{kaas}, to arrive at self-contained asymptotic expressions for $K_{t}$ and $K_{2}$. The mathematical theorem dealing with the angular averages has been discussed in detail in Onsager's original paper \cite{Onsager} and was used later on in Odijk's work on elastic constants \cite{odijkelastic}.  The Gaussian form enables a simple relation $\dot{\psi}_{G} \sim \alpha_{\ell} \psi_{G}  $ to  obtain the derivate of the orientational distributions featuring in \eq{kaas}.  Straightforward algebra  then gives a  simple expression for the helical amplitude:
\beq
K_{t}  \sim  2 \kbt  \bar{\epsilon} (LD)^{2} \rho_{0}^{2}
\label{ktll}
\eeq 
independent of the monomer distribution  as was found for rigid rods with a fixed  length polydispersity \cite{Wensink_2019}.  This result emerges from the simple bilinear dependence of \eq{ommt} on the effective segment length (in this case $\lambda_{\ell}$). The result holds for any  effective length scale $\ell_{k} < \ell$. Dividing each polymer into short sections of length leads to a reduction $M_{t}$ but the loss is exactly compensated by a simultaneous increases of the segment number density.  

The bilinear dependence on segment length is no longer valid for the twist elastic modulus [cf. \eq{omomm2}] which takes on a more complicated form:
\begin{align}
K_{2}     \sim & \frac{\kbt}{96} L^{4} D  \left( \frac{\pi}{2} \right ) ^{\frac{1}{2}}  
\sum_{  \ell ,   \ellp}  \rho_{\ell}   \rho_{ \ellp} \ell \ellp \left ( \frac{1}{\alpha_{\ell} } + \frac{1}{\alpha_{\ellp}} \right )^{\frac{1}{2}}  \nonumber \\ 
& \frac{\lambda_{\ell}^{2} \alpha_{\ellp} (4 \alpha_{\ell} + 3 \alpha_{\ellp} ) + \lambda_{\ellp}^{2} \alpha_{\ell} (3 \alpha_{\ell} + 4 \alpha_{\ellp}) }{( \alpha_{\ell} + \alpha_{\ellp} ) ^{2}}
\label{k2full}
\end{align}
where we reiterate the aggregate-dependent  deflection length $\lambda_{\ell}  = \ell_{p} / \alpha_{\ell}$. In case of a nematic phase of monodisperse, perfectly rigid rods, we substitute $\ell = \ellp = 1$ and $\alpha_{\ell} = \alpha_{\ellp} = \alpha $.  Defining a dimensionless concentration $c_{0} = \rho_{0} v_{0}$ and minimizing the nematic free energy (with $\varepsilon_{b} \rightarrow \infty $ and $q=0$) with respect to $\alpha$ (leading to $\alpha \sim 4c_{0}^{2}/\pi $) one easily recovers  $\beta K_{2}D \sim 7c_{0} /24 \pi$ \cite{odijkelastic}. This simple result suggests a linear increase of the twist elastic constant with rod concentration. The moduli for the splay and bend deformations can be derived in a similar manner and the results are shown in the Appendix.

The equilibrium pitch $p_{\text{chol}}$ of the polymer cholesteric follows from  minimizing \eq{free} with respect to $q$ and reflects a balance between the helical amplitude and twist elastic modulus:
\beq
q \equiv \frac{2 \pi} {p_{\text{chol}}} =  \frac{K_{t}}{K_{2}}
\label{eqpitch}
\eeq
Clearly,  the pitch will be very sensitive to changes in the monomer density $\rho_{\ell}$ as well as the distribution of alignment strength $\alpha_{\ell}$ of the individual monomeric rods. These distributions are intricately coupled and their equilibrium forms follow from an algebraic minimization of the  free energy that we shall address next. 
 
\section{Equilibrium aggregate density}

Using the asymptotic expressions presented above into  \eq{free}, we obtain a much simpler representation for the free energy. For non-chiral monomers ($q=0$) we  find
\begin{align}
 \frac{\beta F^{(N)}v_{0}}{ V} &  \sim   \sum_{ \ell} c_{\ell} \ell^{-1}  \left [ \ln \left (  \ell^{-1}  c_{\ell} \alpha_{\ell} \right ) - 2  - \beta \varepsilon_{b} + \frac{\ell}{4\ell_{p}} \alpha_{\ell}   \right ]  \nonumber \\ 
 & + \beta \varepsilon_{b} c_{0} + \frac{a_{1}}{2}   \sum_ {\ell ,  \ellp}  c_{\ell} c_{\ellp} \left ( \frac{1}{\alpha_{\ell}} +  \frac{1}{\alpha_{\ellp}} \right )^{\frac{1}{2}}   
 \label{free_ast}
\end{align}
with $\beta = 1/\kbt$,  $c_{\ell} =  \rho_{\ell} v_{0}$ a dimensionless aggregate density, and  $a_{1} = (8/\pi)  (\pi/2)^{1/2}$ a constant. Minimizing with respect to $\alpha_{\ell}$ we find that the equilibrium form for $\alpha_{\ell}$ is a solution of the following nonlinear summation equation
\beq
 \alpha^{\frac{1}{2}}_{\ell} + \frac{\ell}{4\ell_{p}} \alpha^{\frac{3}{2}}_{\ell} -  \frac{a_{1}}{2}  \ell  \sum_{  \ellp} c_{\ellp} \left ( 1+ \frac{\alpha_{\ell}}{\alpha_{\ellp}}\right )^{-\frac{1}{2}} =0
 \label{acubic}
\eeq
For infinitely stiff aggregates $\ell_{p} \rightarrow \infty$ the second term vanishes and the condition coincides with the one established for rod nematics with quenched length dispersity \cite{wensink2003,Wensink_2019} were $\alpha_{\ell} $ was found to increase quasi-linearly with $\ell$ for contour lengths larger than the average value. The condition above cannot be solved in closed form, but the distribution $\alpha_{\ell}$  is easily found numerically by combining the real analytical solution of the cubic equation for $\alpha_{\ell}^{1/2}$  with a simple iteration scheme on an equidistant grid of $\ell$ values on the interval $\ell \in [ 1, \ell_{\rm max} ]$ with $\ell_{\rm max} >150$ with the cut-off value depending on the expected degree of polymerization.

If the persistence length is not too large, say $\ell_{p} <10$, the coupling between chain length and nematic order turns out to be very weak and it is generally safe to assume $\alpha_{\ell} $  independent of $\ell$, so that $\alpha_{\ell} \equiv \alpha$. Then, \eq{acubic} is replaced by a simpler algebraic equation: 
\beq
\left (\sum_{ \ell} c_{\ell} \ell^{-1} \right )  \alpha^{\frac{1}{2}} + \frac{c_{0}}{4 \ell_{p}} \alpha^{\frac{3}{2}} - \frac{a_{1}}{4} 2^{\frac{1}{2}} c_{0}^{2} =0
\label{cona}
\eeq 
This equation still needs to be solved self-consistently as the total concentration of aggregates featuring in the first term is unknown {\em a priori}.

The next step is to functionally minimize \eq{free_ast} with respect to the monomer distribution $c_{\ell}$ giving an Euler-Lagrange equation:
\beq
\frac{\partial}{\partial c_{\ell}} \left ( \frac{\beta Fv_{0}}{V} + \lambda \sum_{  \ell} c_{\ell} \right ) =0
\eeq
The Lagrange multiplier $\lambda$ ensures that the monomer concentration be preserved, i.e.,  $\sum_ {\ell} c_{\ell} = c_{0}$.   Let us first focus  on an {\em isotropic} system for which the free energy takes the following simple form:
\begin{align}
\frac{\beta F^{(I)}v_{0}}{ V}  \sim &  \sum_{   \ell} c_{\ell} \ell^{-1} \left [ \ln \left (\ell^{-1} c_{\ell}  \right ) - 1 - \beta \varepsilon_{b} \right ] \nonumber \\ 
& + \beta \varepsilon_{b} c_{0}   + c_{0}^{2}   
 \label{free_iso}
\end{align}
The minimum free energy corresponds to a simple geometric distribution:
\beq
c_{\ell} ^{(I)} = \ell e^{\beta \varepsilon_{b}}  \left ( 1- \frac{1}{m_{0}}  \right )^{\ell}
\label{ri}
\eeq
Recalling  the contour length distribution is given by $c_{\rm a}(\ell ) = \ell^{-1}  c_{\ell}$ we define the average aggregation number in the isotropic phase:
\begin{align}
\langle \ell \rangle  &= \frac{\sum_{ \ell} c_{\rm a} (\ell) \ell}{\sum_{  \ell} c_{\rm a}(\ell)}  = m_{0}     
\end{align}
Conservation of the number of monomers dictates that $m_{0}$ increase with the overall monomer concentration $c_{0}$ and bond energy $-\varepsilon_{b}$ via :
\beq
m_{0} = \frac{1}{2} \left (  1 + \sqrt{1+ 4c_{0} e^{-\beta \varepsilon_{b}} }  \right )
\label{cons}
\eeq
The contour length distribution for the nematic (or cholesteric) phase is much more complicated.  Some algebra leads to a generic exponential form (which is the continuum analog of the geometric distribution for the isotropic case), namely:
\beq
c_{\ell}^{(N)} = \ell e^{\beta \varepsilon_{b}}  \exp \left (-\frac{\ell } {  m_{\ell} } \right )
\label{rhon} 
\eeq
where the effective mean aggregation number $m_{\ell}$ is a non-analytical function of $\ell$:
\begin{align}
\frac{1}{m_{\ell}}  = &   \lambda + \frac{1} {\ell}  (\ln \alpha_{\ell} -1) + \frac{1}{4 \ell_{p}} \alpha_{\ell} + a_{1} \int d \ellp c_{\ellp} \left ( \frac{1}{\alpha_{\ell}} +  \frac{1}{\alpha_{\ellp}} \right ) ^{\frac{1}{2}} 
\label{malpha}
\end{align}
The intrinsic coupling between $c_{\ell}$ and $\alpha_{\ell}$  necessitates \eq{rhon} and (\ref{malpha})  be solved self-consistently in conjunction with \eq{acubic} to yield the equilibrium aggregate density $c_{\ell}$ and distribution of the nematic alignment parameter  $\alpha_{\ell}$ for a given overall monomer concentration $c_{0}$, persistence length $\ell_{p}$ and bond energy $\varepsilon_{b} $. 

Coexistence between isotropic and nematic (cholesteric) phases can be established from equating the chemical potential and pressure in each phase.  For the nematic phase we obtain a chemical potential  ($  \mu  \equiv  \partial (Fv_{0}/V) / \partial c_{0}$) ignoring constant factors that are identical to both phases:
\begin{align} 
 \beta \mu^{(N)}  & \sim   \frac{1}{c_{0}} \sum_{\ell}   c_{\ell}  \ell^{-1}  \left [ \ln \left ( \ell^{-1}   c_{\ell} \alpha_{\ell} \right ) - 1  - \beta \varepsilon_{b}  + \frac{\ell}{4 \ell_{p}} \alpha_{\ell}   \right ] \nonumber \\
 & + \beta \varepsilon_{b} +  \frac{ a_{1} }{c_{0}}  \sum_{  \ell , \ellp}  c_{\ell} c_{\ellp} \left ( \frac{1}{\alpha_{\ell}} +  \frac{1}{\alpha_{\ellp}} \right ) ^{\frac{1}{2}} 
\end{align}
The osmotic pressure $P= (F- N\mu)/V$ combines an ideal gas term with an excluded-volume contribution:
\begin{align}
 \beta P^{(N)} v_{0} & \sim \sum_ {\ell} c_{\ell} \ell^{-1}  +   \frac{a_{1}}{2} \sum_{  \ell , \ellp }  c_{\ell} c_{\ellp} \left ( \frac{1}{\alpha_{\ell}} +  \frac{1}{\alpha_{\ellp}} \right ) ^{\frac{1}{2}} 
\end{align}
Likewise for the isotropic phase, we obtain the following simple expressions from \eq{ri} and \eq{cons} :
\begin{align} 
\beta \mu^{(I)}  & \sim \ln \left (  1 - \frac{1}{m_{0}} \right ) + \beta \varepsilon_{b}  + 2 c_{0} \nonumber \\ 
\beta P^{(I)} v_{0} & \sim e^{\beta \varepsilon_{b}} (m_{0} -1 ) + c_{0}^{2}
\end{align}
The  three relevant parameters controlling  isotropic-nematic coexistence are the overall monomer concentration $c_{0}$,  the  effective temperature $\varepsilon_{b}$  that quantifies the propensity for the rod tips to bond together, and the persistence length $\ell_{p}$ with $\ell_{p} \rightarrow \infty$ corresponding to the limiting case of completely rigid polymers. The conventional isotropic-nematic transition of unaggregated  rods is retrieved by taking the limit $\beta \varepsilon_{b}  \rightarrow \infty $ in which case the rod tips should be infinitely repulsive.  The mean aggregation number $m_{0} \approx 1+ c_{0} e^{-\beta \varepsilon_{b}}$  is then asymptotically close to unity and the pressure ($\beta P^{(I)} v_{0} \sim c_{0} + c_{0}^{2}$) and chemical potential ($\beta \mu^{(I)}   \sim \ln c_{0} + 2 c_{0} $) are simply those of an isotropic assembly of strictly monomeric rods treated within the second-virial approximation.  Solving the coexistence conditions requires a numerical scheme involving three nested iteration loops; an  inner one solves the nematic order parameter $\alpha_{\ell}$ [cf. \eq{acubic}] for a given aggregate length distribution $c_{\ell}$ and monomer concentration $c_{0}$. This loop is enveloped into a second iteration determining the correct normalization factor $\lambda$ [cf. \eq{malpha}]. Finally,  in the outer iteration loop  the monomer concentrations for the coexisting phases are resolved from the chemical and mechanical equilibrium conditions ($\mu^{(N)} = \mu^{(I)}$ and $P^{(N)} = P^{(I)}$).

\section{Effect of polymerization inhibitors}

Let us now consider the presence of small molecules or nanoparticles that {\em irreversibly} bind to the rod tips and stop the aggregation process.  Once such a particle is bound to either tip of the chain is no longer able to grow at that end. The inhibitors are assumed to be infinitesimally small  such that, when attached to the tips, they do not influence the interaction between the aggregates.  We also assume that the inhibitors do not impart any depletion effect which may lead to effective attraction between the chain segments.  
A further assumption is that the point inhibitors diffuse ultrafast such that thermodynamics, not kinetics is the determining factor of the inhibition process.
The effect of the polymer inhibitors can be described, albeit in a crude manner, in terms of an additional free energy representing a simple entropic potential. Let us propose the following form:
\beq
 \frac{ \beta F_{i} v_{0}}{ V}  \sim \epsilon_{i}  \left \{  \tanh \left [ \sigma \left (c_{i} - c_{\text{tip}}  \right ) \right ] -1 \right \} 
 \label{penalty}
\eeq
with $c_{\text{tip}} = 2 \sum_{\ell} c_{\ell} \ell^{-1}$ the concentration of aggregate tips. The constant $\sigma $ should be much larger than unity mimicking a step function. 
Here, we fix $\epsilon_{i} = 10 $ and $\sigma = 10$ and vary only the inhibitor concentration.
If the inhibitor concentration is larger than the tip end concentration, then there is a free energy penalty defined by the binding propensity $\epsilon_{i} > 0$ of the inhibitors. In the opposite case (tip concentration exceeding the inhibitor concentration) the  penalty is zero and the inhibitors do not interfere with the polymerization process. The additional free energy term modifies the self-consistent expression for the mean aggregation number in the nematic phase \eq{malpha}. It now becomes:
\begin{align}
\frac{1}{m_{\ell}}  = &   \lambda + \frac{1} {\ell}  (\ln \alpha_{\ell} -1) + \frac{1}{4 \ell_{p}} \alpha_{\ell} + a_{1} \int d \ellp c_{\ellp} \left ( \frac{1}{\alpha_{\ell}} +  \frac{1}{\alpha_{\ellp}} \right ) ^{\frac{1}{2}} \nonumber \\ 
& - \frac{2\epsilon_{i} \sigma}{\ell} \sech^{2}  \left [ \sigma \left ( \rho_{i} - 2 \sum_{\ell} \rho_{\ell} \ell^{-1} \right ) \right ]
\label{malpha_inh}
\end{align}
For the isotropic phase, the presence of inhibitors violates the simple  geometric aggregate length distribution and we must consider a  self-consistently form via $c_{\ell}^{(I)} = \ell e^{\beta \varepsilon_{b}}  \exp  (-\ell /  m_{\ell}^{(I)}  )$  using:
\begin{align}
\frac{1}{m_{\ell}^{(I)}}  = &   \lambda  - \frac{2\epsilon_{i} \sigma}{\ell} \sech^{2} \left   [ \sigma \left (\rho_{i} - 2 \sum_{\ell} \rho_{\ell} \ell^{-1} \right ) \right ]
\label{malpha}
\end{align}
with $\lambda$ to be resolved from the imposed concentration of rod monomers $c_{0}$.

\begin{figure}
\begin{center}
\includegraphics[width=  \columnwidth]{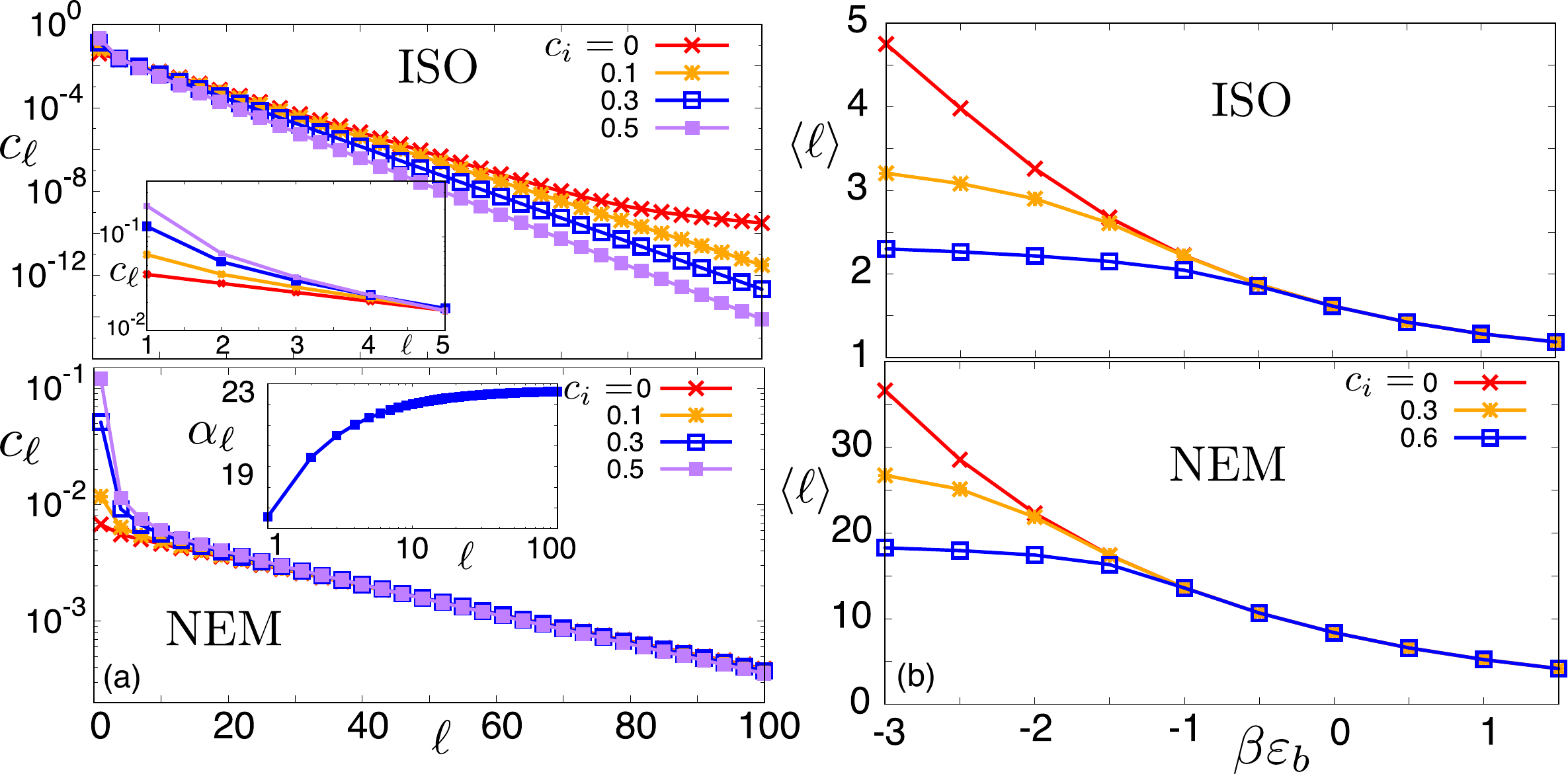}
\caption{ \label{fig2} (a) Contour length distribution of reversibly aggregating  flexible rods in the isotropic phase (at overall monomer concentration $c_{0} =1$) and  nematic phase ($c_{0} =8$) at effective temperature $ \varepsilon_{b} = -3 \kbt$. Inset: Gaussian variational parameter $\alpha_{\ell}$ measuring the degree of nematic order of each aggregate  $\ell$. The  persistence length in units rod length is $\ell_{p} = 3$.   (b) Mean aggregation number $\langle \ell \rangle$ plotted versus the effective temperature $\varepsilon_{b}$. At low temperature, reversible polymerization is strongly hampered by the presence of inhibitors ($c_{i} >0$) leading to shorter aggregates. }
\end{center}
\end{figure}

\begin{figure}
\begin{center}
\includegraphics[width=   \columnwidth]{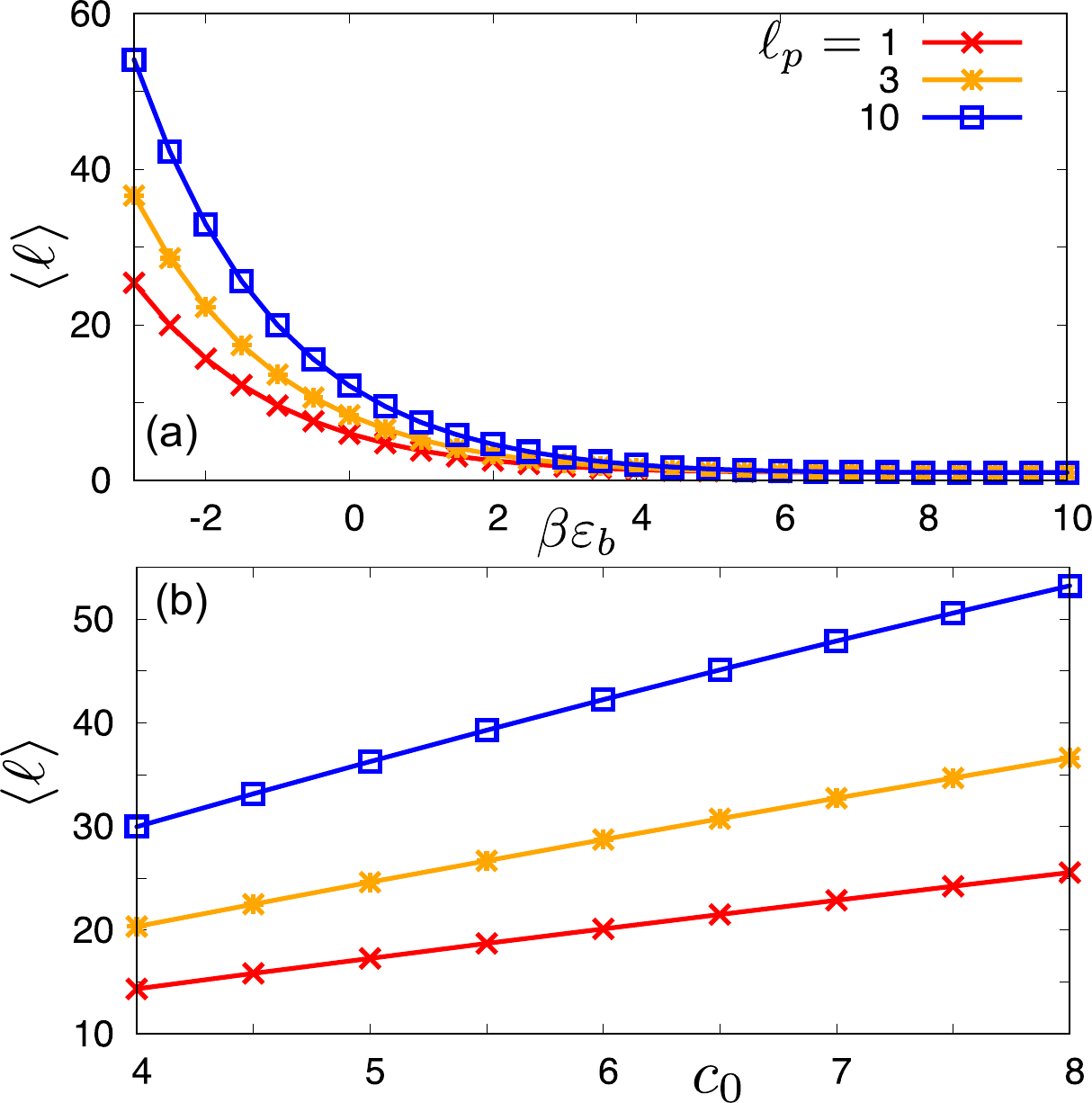}
\caption{ \label{fig3}  (a) Impact of the effective temperature $ \varepsilon_{b}$ on the mean aggregation number $\langle \ell \rangle$ for a polymeric nematic  with overall monomer concentration $c_{0} =8$ explored at various degrees of monomer flexibility expressed by the persistence length $\ell_{p}$. (b) Effect of monomer concentration on the mean aggregation number. In all cases, polymerization is uninhibited ($c_{i} = 0$). }
\end{center}
\end{figure}

The equilibrium aggregate distribution corresponding to a disordered, isotropic fluid and a nematic fluid are depicted in \fig{fig2}.  The  typical bi-exponential form found  in chromonic nematics  \cite{lu2004,kuriabova2010} is far less noticeable here since even the shortest aggregates are strongly aligned as we can see from $\alpha_{\ell}$ at small $\ell$. The effect of adding polymerization inhibitors is as expected; the partial concentration of monomers is strongly enhanced while the probability of encountering long aggregates is reduced. 

For the nematic phase, we observe that the inhibitors primarily increase the statistical weight of monomeric and oligomeric species without too much affecting the probability of the longer species. The  length-dependent nematic order parameter $S_{\ell} \sim 1 - 3\alpha_{\ell} ^{-1} $ turns out only marginally influenced by the concentration of inhibitors. This is likely due to a loss of large aggregates with a strong aligning potential  being offset by a simultaneous gain in mono- and oligomeric species which, contrary to the low anisotropy discotic aggregates found in chromonics, also have a propensity to align. The mean aggregation reveals that polymerization inhibition, expressed by a reduction of the mean aggregation number $\langle \ell \rangle$, is only noticeable at low  temperature $\varepsilon_{b}$.  Beyond a critical  temperature the number concentration of free ends surpasses $c_{i}$ so that all inhibitors 
can bind without the need to enforce fragmentation of the polymers in order to generate more free ends. 

\begin{figure}
\begin{center}
\includegraphics[width=   \columnwidth]{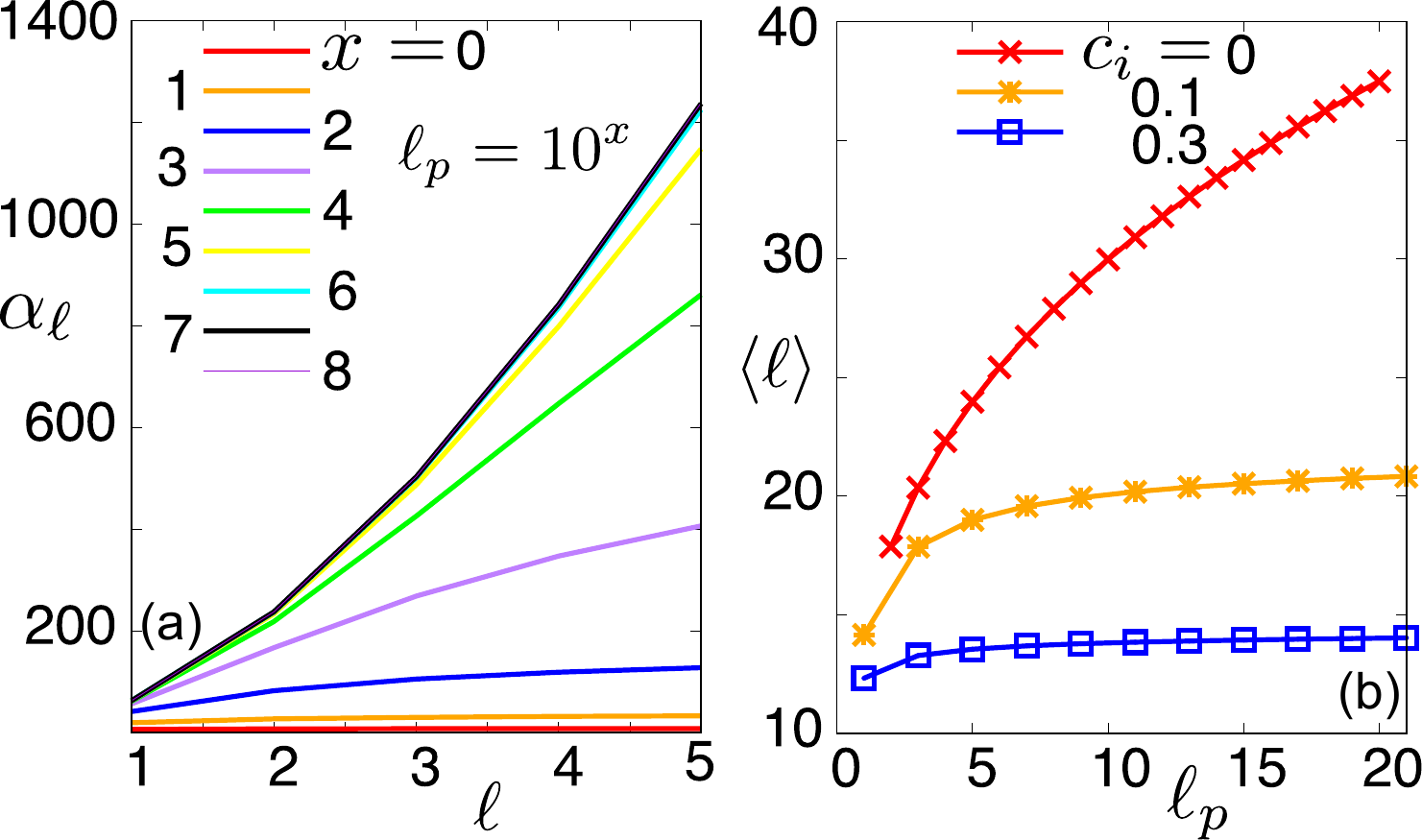}
\caption{ \label{fig7} (a) Nematic order for small aggregates with a {\em fixed} length distribution \eq{ri} with ($c_{0}=5$, $ \beta \varepsilon_{b} = -3$) at different persistence length $\ell_{p}$. The black curve corresponds to the result for perfectly rigid rods ($ \ell_{p} \rightarrow \infty $). (b) Impact of polymerization inhibitors on the average aggregate length for increasingly stiffer monomers.    }
\end{center}
\end{figure}

The effect of  flexibility on the aggregation behaviour  is demonstrated in \fig{fig3}. As expected, the mean aggregation number is higher for stiff rods than for the more flexible ones at comparable temperature.  Strictly, one would expect the rods to grow infinitely long in the physically unrealistic limit of perfectly rigid rods $\ell_{p} \rightarrow \infty$   \cite{vdschoot1994la,vdschoot1994epl}.  \fig{fig7}a demonstrates the evolution of the size-dependent nematic order for  persistence lengths across various orders of magnitude.  To avoid numerical complications involving extraordinary long polymers formed by increasingly stiff rods, a  fixed (i.e. quenched) length distribution was adopted here. Clearly, the rigid rod limit is only reached at extremely large persistence lengths, typically $\ell_{p} > 10^{6} $, way beyond the values considered in the calculations.

\fig{fig7}b demonstrates that the aggregate growth  for increasingly rigid rods is strongly curbed by adding inhibitors, suggesting a finite average aggregation number may be reachable in the stiff rod limit provided the inhibitor concentration is not too small (in this case $c_{i} > 0.1$).

We reiterate that the  Ansatz \eq{penalty} is purely intuitive and should capture the main consequences of irreversibly binding inhibitors.  The case of  a {\em finite}  binding affinity, say of the order of the $\varepsilon_{b}$, requires a more elaborate theory  involving two coupled segment densities (one for capped and another one for uncapped aggregates) that would follow self-consistently from minimizing the free energy with respect to these densities.  We refrain from such a theory here in view of technical complications this would involve.

%%% LATEST FIGURE

\section{Isotropic-nematic phase diagrams}

\begin{figure}
\begin{center}
\includegraphics[width=   \columnwidth]{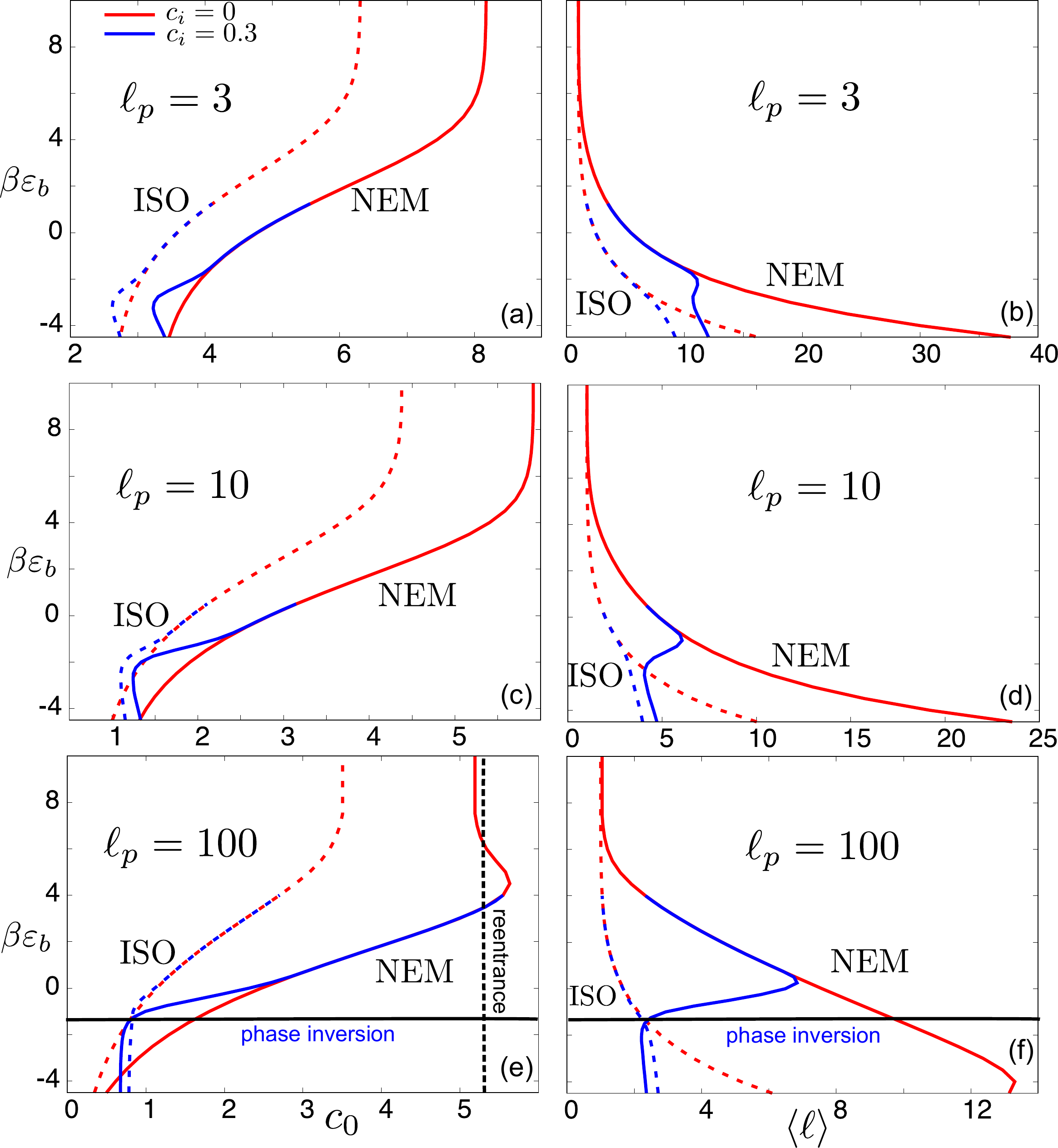}
\caption{ \label{fig4} Overview of the isotropic-nematic  phase diagrams for reversibly polymerizing weakly flexible rods at various persistence length $\ell_{p}$. Binodals  are represented in terms of the overall monomer concentration $c_{0}$ (left panels) and mean aggregation number $\langle \ell \rangle$ (right panels) as a  function of the effective temperature $ \varepsilon_{b}$. The isotropic branches are dotted, the nematic ones are shown as  solid lines.  At high temperature $\beta \epsilon_{b} \gg 1$ polymerization is strongly suppressed and system is principally composed of monomeric rods so that $\langle \ell \rangle \approx 1$ for both coexisting phases.  The presence of polymerization inhibitors $c_{i} >0$ (blue curves) strongly alters the phase behavior at low temperature introducing marked reentrance effects (vertical dotted black line) and isotropic-nematic phase inversions (indicated by the region below the horizontal black line).   }
\end{center}
\end{figure}

Beyond a critical monomer concentration the isotropic fluid becomes unstable with respect to nematic order.   Unlike in previous models for chromonics and related systems \cite{lu2004,kuriabova2010,nguyen2014,romani2018} where the monomer shape is insufficiently anisotropic to guarantee stable nematic order, the filamentous monomers considered here exhibit stable nematic order even when unpolymerized ($\beta \varepsilon_{b} \gg 1$). The impact of the effective temperature controlling the degree of polymerization is demonstrated in \fig{fig4}a and b. Focussing first on the inhibitor-free systems ($c_{i} =0$) we observe that, at low temperature,  polymerization increases the average chain length of the aggregates thereby enhancing the propensity of the system to form a  nematic phase. As  a  result, the binodals shift to lower $c_{0}$ upon decreasing $\varepsilon_{b}$.  The same mechanism albeit at a more extreme level is at play for stiffer monomers with $\ell_{p} = 10$ (\fig{fig4}c and d).  Somewhat contrary to the trend in \fig{fig3}a, we observe that at fixed temperature the average chain length in the nematic phase becomes  {\em smaller} than for the more flexible monomers in \fig{fig4}b. This effect could be ascribed to the fact that the monomer concentration of the coexisting nematic phase strongly decreases with persistence length, suggesting the effect of monomer crowding to be more determining for the mean aggregation number than the monomer flexibility.  For near-rigid monomeric rods ($\ell_{p} =100$)  a reentrant phase transition is observed involving a sequence of states  N $\rightarrow $  I+N  $\rightarrow $ N upon lowering the temperature. The average chain length in the nematic phase at coexistence has dropped even further compared to the flexible monomers. We reiterate that all numerical results presented in this work  respect the implicit chain-length distribution of the nematic order parameter through the distribution $\alpha_{\ell}$ [cf. \eq{acubic}].  For all data points explored thus far we found that the nematic alignment of the  rod segments  is sufficiently strong ($\alpha_{\ell=1}  > 10$) to ensure the Gaussian parameterization \eq{gauss} to be valid.

Let us now turn to analyzing the effect of the inhibitors.  As noted before, their influence is only defined in the low temperature regime (strong polymerization)  where the concentration of free ends is at par with the inhibitor concentration. The presence of the inhibitors  drastically modifies  the course of the binodals at low temperature, introducing a marked reentrance effect in the isotropic phase (giving a complex sequence of phases; I $\rightarrow $ I+N $\rightarrow$ N $\rightarrow $ I+N upon lowering $ \varepsilon_{b}$). The reentrance effect is accompanied by a strong decrease in the average aggregate length in the nematic phase, particularly for the near-rigid rods ($\ell_{p} = 100$).  

 The nose-shaped reentrant sections in \fig{fig4}b,d and f, where the average aggregate length decreases with lowering temperature, are somewhat counterintuitive but can be rationalized in terms of polymers being fragmented by the presence of inhibitors. The reentrance sections in \fig{fig4}a and b suggest that a uniform nematic phase may enter isotropic-nematic coexistence upon lowering temperature. Most likely,  this is a simple consequence of the fact that the shortened aggregates prefer to order isotropically (because their excluded-volume entropy penalty is less severe than for long polymers) as inhibitor-driven fragmentation occurs.
The qualitative features of the phase diagrams are robust against changing the values of $\epsilon_{i}$ and $\sigma $  in the  penalty contribution \eq{penalty}. However, the stability of the numerical calculations is compromized for very large values ($ \gg 10$).

For the stiffest rods ($\ell_{p} =100$), a curious phase inversion is observed below a certain effective temperature. In this regime, the monomer concentration as well as the mean aggregation number of the isotropic phase  are slightly {\em higher} than those of the coexisting nematic phase. This subtle effect is unseen in monodisperse systems but has been observed in binary colloidal mixtures  \cite{wensink2001,schmidt2004} and is entirely driven by the inherent size-dispersity of the aggregates. The phase inversion scenario may have interesting consequences for the phase-separation process involving nematic tactoids with a mass density similar to that of the isotropic fluid in which they are embedded as found for thermotropic liquid crystals.  

In our analysis, we have not included the columnar phase which will most likely appear as a stable phase at large monomer concentrations \cite{schoot1996,kuriabova2010}.  For slender monomeric rods (say $L/D \sim 100$) stable columnar order would be expected at fairly elevated monomer concentrations $c_{0} > \phi_{0} (L/D) $ (taking a typical critical packing fraction $\phi_{0} \sim 0.4 $ associated with the onset of columnar order). The possibility of columnar ordering is therefore unlikely to interfere with the fluid phase diagrams presented in \fig{fig4}.   The possibility of liquid-liquid phase separation involving two isotropic or nematic phases can easily be verified from the pressure and chemical potential curves but no such  transition was found at least for the parameter ranges explored in this study. This suggests that aggregate length disparity alone is not enough to drive liquid-liquid phase separation (as found for rods with quenched length-dispersity \cite{speranza2003,wensink2003}) and that interchain attractions are a necessary ingredient \cite{kindt2001}.

\section{Results for the elastic moduli and pitch}

\begin{figure}
\begin{center}
\includegraphics[width=   \columnwidth]{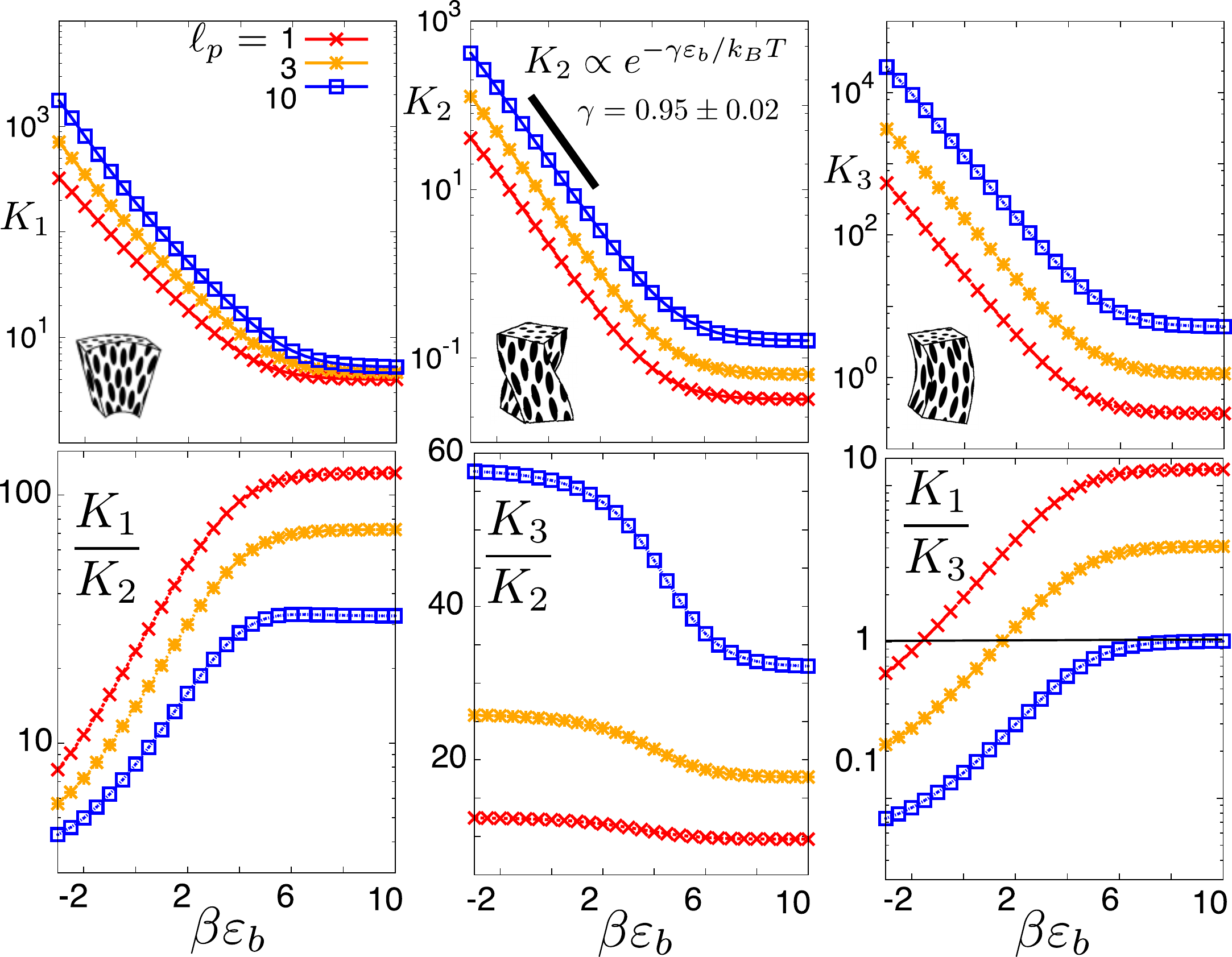}
\caption{ \label{fig5}   Temperature-dependence of the splay ($K_{1}$), twist ($K_{2}$) and bend ($K_{3}$) elastic moduli (in units $k_{B}T/D$) for a nematic fluid of reversibly polymerizing rods with overall monomer concentration $c_{0} = 8$. The bottom panels depict the elastic anisotropy in terms of  the splay-twist, bend-twist, and the  splay-bend elastic ratio.  }
\end{center}
\end{figure}

At low temperature, the rod tips are strongly attractive and favor strong polymerization leading to increasingly longer filaments.  This is accompanied by a dramatic increase of the elastic moduli (\fig{fig5}c). Since the helical amplitude is constant with $c_{0}$ and the molecular chirality $\bar{\epsilon}$ is only weakly dependent on temperature, the behavior of the cholesteric pitch is largely enslaved to changes in $K_{2}$. A heuristic fit of the data above suggests an exponential change of the pitch with effective temperature via:
\beq
\frac{p_{\text{chol}}}{L} \propto \exp \left ( - \gamma \frac{  \varepsilon_{b}}{\kbt}  \right ) 
\eeq
Upon fitting the curves we find  a value close to unity $\gamma =  0.95 \pm 0.02 $. Varying the monomer concentrations $c_{0}$ within the stable nematic regime produces values that are within the error margin. The exponential relationship thus reveals a marked temperature-response of the pitch for cholesterics of reversibly polymerizing chiral rods, much stronger than what  is commonly observed in most monomeric chiral liquid crystals \cite {hiltrop,dierking2014} and chromonics where a pitch increase with temperature was recently observed \cite{ogolla2019}. In fact, the strong increase in the twist elasticity in the polymeric regime is at odds with chromonic liquid crystals where the twist modulus is found to be very small compared to the splay and bend moduli \cite{zhou2012,romani2018}. The weak twist facilitates the emergence of spontaneously twisted patterns in cylindrical confinement \cite{nayani2015,jeong2015,prinsen2004}. Our findings therefore suggest that a spontaneous twist of the director field aided by a specific system geometry  is less likely to occur in rod-based polymeric nematics.

\begin{figure}
\begin{center}
\includegraphics[width=   \columnwidth]{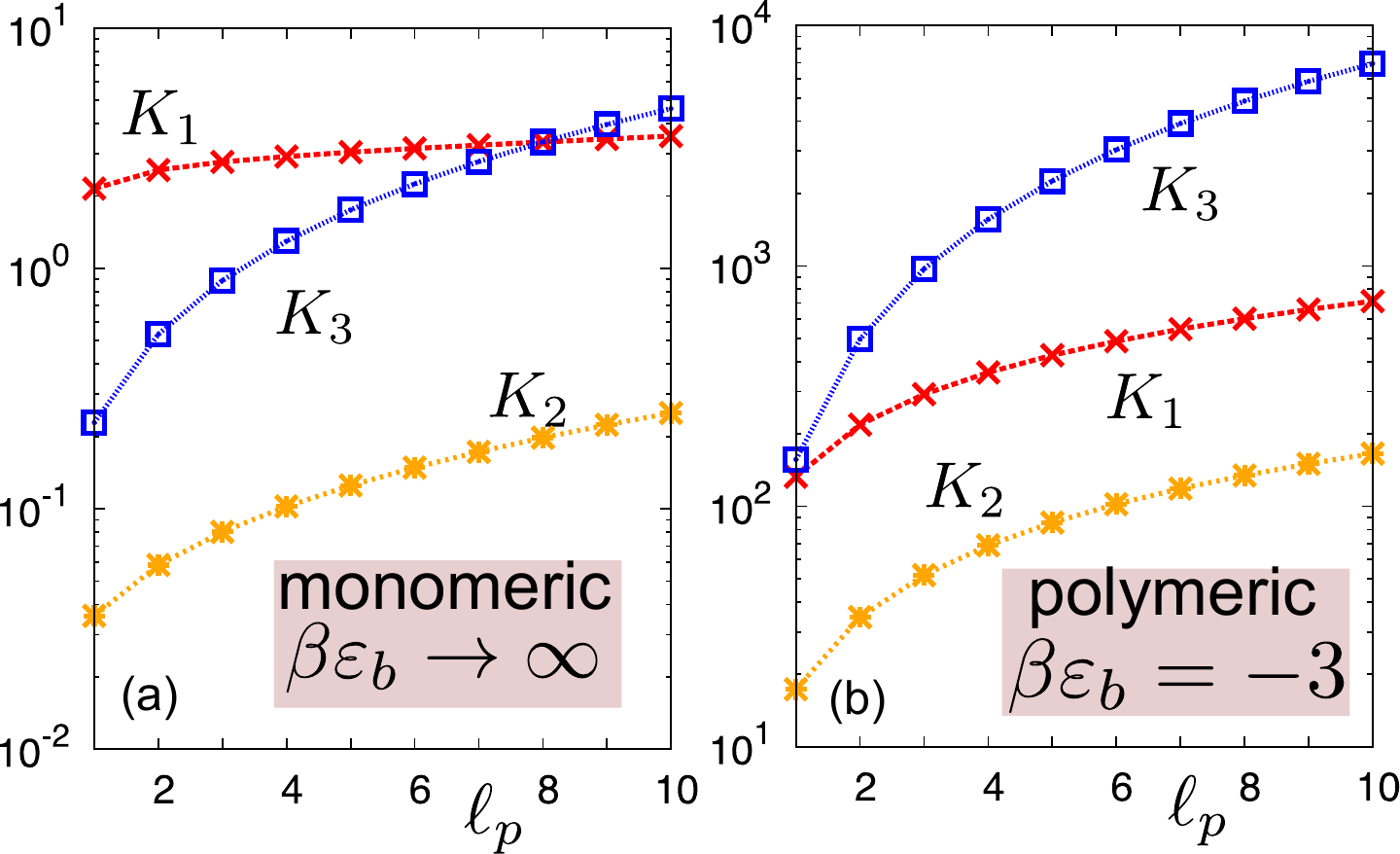}
\caption{ \label{fig6}  Impact of the  persistence length $\ell_{p}$ on the elastic moduli for a (a) monomeric nematic  at infinite temperature, and (b) a polymeric nematic  at $\varepsilon_{b} = - 3 k_{B}T$. In both cases $c_{0} = 4$.  }
\end{center}
\end{figure}

\begin{table*}
\caption{\label{table1}
Overview of the elastic anisotropies of several main classes of lyotropic nematics in terms of the principle (${\bf I}$), intermediate (${\bf II}$) and minor (${\bf III}$) elastic modulus.}
\begin{tabular}{l c c c}
    \hline
{\bf system}& {\bf I} & {\bf II} & {\bf III}  \\
weakly flexible monomeric rods ($1<\ell_{p} \leq 10$) \cite{gemunden2015}${\text{[\fig{fig6}]}}$  & \red{splay} & \blue{bend}  & \amber{twist}  \\
rigid monomeric rods ($\ell_{p} > 10$) \cite{odijkelastic,milchev2018} &  \blue{bend} & \red{splay}  & \amber{twist} \\
rigid monomeric disks \cite{sokalski1982,obrien2011,wensink2018}  & \amber{twist} & \red{splay} & \blue{bend} \\
polymeric disks (chromonics) \cite{zhou2012,romani2018}  & \red{splay} & \blue{bend} & \amber{twist} \\  
polymeric rods${\text{ [\fig{fig6}]}}$ & \blue{bend} & \red{splay} & \amber{twist} \\
    \hline
\end{tabular}
\end{table*}

An overview of the elasticity anisotropies is given in  \fig{fig5}. The results showcase a dramatic change in elastic properties when going from a monomeric nematic at high temperature to a highly length-disperse aligned polymeric nematic at low temperature.  
We observe a marked drop in the splay-twist and splay-bend ratio with decreasing temperature which is  opposite to what has been recently reported from a density-functional theory for chromonic liquid crystals \cite{romani2018}. This suggests that the elastic response of a rod-based polymeric nematic is qualitatively different from chromonics where the three elastic anisotropies  were  all found to enhance with decreasing temperature. 
We attribute the discrepancy with chromonics to the fact that those systems tend to be characterized by a large fraction of rigid, low-anisometry short-fragments that are randomly oriented within a nematic matrix \cite{romani2018}  whereas  the shortest fragments in our systems are strongly anisotropic and co-align with the long chains.

For  weakly flexible rods ($1 < \ell_{p} \leq 10$)  we observe that  while splay dominates the bend elasticity for monomeric nematic  the opposite case (bend elasticity stronger than splay) is found  for the polymeric nematic. In all cases,  the twist modulus turns out to be the minor  contribution. Increasing the stiffness of the rod monomers  leads to anomalous increse of the bend modulus which becomes the principle modulus for rigid  rods ($\ell_{p} \gg 10$).  We remark that in view of the relatively small persistence lengths probed, the use of a constant $\alpha$ approximation (cf. \eq{cona}) should  be entirely justified in  this regime.  Curves obtained from  \eq{acubic}  and \eq{cona} in \fig{fig5} were found to virtually overlap.  
Lastly, we remark that although the elastic moduli  vary  with the monomer concentration and  the inhibitor concentration (results not shown here), no major qualitatively changes in the elastic anisotropies were observed upon changing these variables.

\section{Conclusions}

Slender,  rod-shaped colloidal rods carrying attractive sites at either tip end may associate into linear aggregates. At elevated monomer concentration the average length of the aggregates is sufficient to generate stable nematic or cholesteric phases (in case the monomeric rods are chiral).   
Building on previous modelling efforts \cite{vdschoot1994la,vdschoot1994epl,kuriabova2010,nguyen2014} we have developed a second-virial  theory describing such assemblies, their phase behavior and nemato-elasticity in terms of the monomer concentration, degree of flexibility and effective temperature controlling their propensity to polymerize.  The theory explicitly accounts for chain-length dependence of nematic order through the use of a simple Gaussian representation of the orientation distribution of the monomers.  We find that polymerization drastically alters the elastic properties of the nematic fluid, most notably the twist elasticity which increases several orders of magnitude over a relative narrow temperature range. Our findings underscore the fundamental difference in elastic anisotropy between a monomeric nematic fluid at high temperature and polymeric one at low temperature. The elastic anisotropies in the monomeric nematic regime turn out to be qualitatively similar to those of a chromonic liquid crystal with the twist modulus being the weakest followed by the bend and splay modes ($K_{1} > K_{3} > K_{2}$). A polymeric rod-based nematics on the other hand is dominated by bend elasticity with splay being of intermediate strength  ($K_{3} > K_{1} > K_{2}$).  In Table I  we present an overview of the principal, intermediate and minor elastic moduli for different classed of lyotropic nematics explored thus far in literature. 
The particular elastic anisometry of the nematic material  has important consequences for the shape and mesostructure of nematic droplets \cite{prinsen2003,debenedictis2016} and for their role as an embedding medium in case of colloidal inclusions with interactions mediated by topological distortions of the director field \cite{smalyukh2018,Mundoor18}.  

We have also addressed the effect of polymerization inhibitors that irreversibly bind to the tip end of the filaments thereby terminating the polymerization process. The inhibition effect is taken into account through the use of a simple free energy penalty and its consequences for the phase behavior and elastic properties are systematically investigated. The inhibitors strongly modify the low-temperature part of the isotropic-nematic phase diagram by generating a remarkable reentrance effect as well as a dramatic reduction of the isotropic-nematic phase gap including a phase inversion in the case of very stiff aggregates. In the latter case, a counterintuitive situation is observed where the isotropic phase fraction has both a higher  particle concentration and mean aggregation number than the coexisting nematic phase.  This suggests the phase coexistence to be the result of a very delicate trade-off between orientation, packing and mixing entropy compounded with the association energy featuring in \eq{free}. 

We underscore that the present theory should serve as a mere qualitative assessment of the main trends imparted by polymerization, directionality and flexibility on the fluid phase behavior and nematic elasticity of supramolecular filaments.  Quantitative predictions for the nematic elastic constants cannot be expected from the present scaling theory \cite{romani2018}. In fact, recent advances in modelling semiflexible polymers suggest that
state-dependent effective shape of the polymer \cite{dennison2011} as well as coupled density-orientation fluctuations \cite{egorov2016,milchev2018} may have  important consequences for the collective properties of semiflexible polymers (with persistence length smaller than the contour length). Further simulation work is clearly needed to arrive at a  quantitatively reliable prediction of the nemato-elastic properties and phase behavior of supramolecular assemblies of semi-flexible rods. Nevertheless, the tractability and computational ease of the present theory will hopefully be instructive  in guiding experimental work on confined chromonics \cite{nayani2015,jeong2015} and other reversibly polymeric nematics where knowledge of the elastic anisotropy (along with the surface elastic moduli that we did not address here) is indispensable.

\subsection*{Appendix: Splay and bend elasticity}

For completeness  we present  scaling expressions for the other two elastic moduli related to  splay ($K_{1}$) and bend ($K_{3}$) deformations of the director field. These can be obtained using an  asymptotic analysis similar to the one performed for the twist modulus \cite{odijkelastic}. The resulting expressions read:
\begin{align}
K_{1}  \sim & 3 K_{2} + \Delta K_{1}  \nonumber \\
K_{3}    \sim &  \frac{\kbt}{48} L^{4} D  \left( \frac{\pi}{2} \right ) ^{\frac{1}{2}}  
\sum_ { \ell  , \ellp}  \rho_{\ell}   \rho_{ \ellp} \ell \ellp (\lambda_{\ell}^{2} + \lambda_{\ellp}^{2}) \alpha_{\ell} \alpha_{\ellp}  \nonumber \\ 
& \times  \left [ 3 \left ( \frac{\alpha_{\ell}  +\alpha_{\ellp} }{\alpha_{\ell} \alpha_{\ellp} } \right )^{\frac{3}{2}}  - \frac{3 \alpha_{\ell}^{2} + 4 \alpha_{\ell} \alpha_{\ellp} + 3 \alpha_{\ellp}^{2}}{\alpha_{\ell}^{\frac{3}{2}} \alpha_{\ellp}^{\frac{3}{2}} (\alpha_{\ell} + \alpha_{\ellp})^{\frac{1}{2}}} \right ] 
\end{align}  
The  constant ratio between the splay and twist  moduli is identical to that of monodisperse rigid rods and is unaffected by length dispersity.    We also deduce that the bend contribution for a monodisperse nematic of perfectly rigid hard rods obeys $\beta K_{3} D \sim 4 c_{0}^{3}/3 \pi^{2}$ increasing with the cube of the rod concentration \cite{odijkelastic}. 
For the splay mode,  a correction $\Delta K_{1} $  must be added arising  from density gradients  that are inherently generated by a splay deformation of the director. A non-uniform aggregate density  is compensated by  a gradient in the local concentration of free ends for which an entropic penalty holds. This effect has been discussed by Meyer \cite{meyerbook,taratuta1988} and later corroborated by Kamien et al.  \cite{kamien_nelson1992}.  The penalty term is proportional to the average  contour length $ l$ projected onto the $z-$axis times the areal density $ \rho_{z, \ell} = N_{\ell} l /V  = \rho_{\ell} l\ell^{-1}$ cutting a constant planar cross section along the director. For a polydisperse nematic we write:
\beq
\Delta K_{1} \sim   \frac{1}{2} \kbt \sum_{\ell} \rho_{\ell} \ell^{-1} l^{2}
\eeq
 The projected polymer length is approximated as $ l \approx \ell L \langle \cos \theta \rangle \sim \ell L (1- \alpha_{\ell}^{-1} ) $  so that:
\beq
\Delta K_{1} \sim  \frac{2}{\pi} \frac{ \kbt }{D} \sum_{\ell} \ell c_{\ell} (1 - 2 \alpha_{\ell}^{-1} )
\eeq
Generically, we find that the entropic penalty $\Delta K_{1}$ is at least of the same order as  $\sim 3 K_{2}$ and tends to strongly dominate the excluded-volume contribution, in particular at high  temperature (monomeric regime).

\bibliography{pub}

\end{document}